\documentclass[prl,twocolumn,superscriptaddress,showpacs]{revtex4}
\usepackage{amsmath,amssymb,amsfonts,bbm,graphicx}
\newcommand{\ket}[1]{\vert #1 \rangle} \newcommand{\bra}[1]{\langle #1 \vert}

\newcommand{\average}[1]{\left \langle #1  \right\rangle}
\newcommand{\be}{\begin{equation}}
\newcommand{\ee}{\end{equation}}
\newcommand{\bae}{\begin{eqnarray}} \newcommand{\eae}{\end{eqnarray}}

\def\({\left(} \def\){\right)} \def\[{\left[} \def\]{\right]}

\begin{document}
\title{Experimental estimation of entanglement at the quantum limit}
\author{Giorgio Brida} 
\affiliation{INRIM, I-10135, Torino, Italy}
\author{Ivo Degiovanni}
\affiliation{INRIM, I-10135, Torino, Italy}
\author{Angela Florio}
\affiliation{INRIM, I-10135, Torino, Italy}
\affiliation{Dipartimento di Fisica, Universit\`a di Bari, I-70126 Bari, Italy}
\author{Marco Genovese}
\affiliation{INRIM, I-10135, Torino, Italy}
\author{Paolo Giorda}
\affiliation{ISI Foundation, I-10133 Torino, Italy}
\author{Alice Meda}
\affiliation{INRIM, I-10135, Torino, Italy}
\author{Matteo G. A. Paris}
\affiliation{Dipartimento di Fisica, Universit\`a degli Studi 
di Milano, I-20133 Milano, Italy}
\affiliation{ISI Foundation, I-10133 Torino, Italy}
\affiliation{CNISM, UdR Milano, I-20133 Milano, Italy}
\author{Alex Shurupov}
\affiliation{Dipartimento di Fisica, Politecnico di Torino, I-10129
Torino, Italy}
\affiliation{INRIM, I-10135, Torino, Italy}
\affiliation{Faculty of Physics, Moscow State University, 
119992, Moscow, Russia}
\date{\today}
\begin{abstract}
Entanglement is the central resource of quantum information
processing and the precise characterization of entangled states
is a crucial issue for the development of quantum technologies.
This leads to the necessity of a precise, experimental feasible
measure of entanglement.  Nevertheless, such measurements are
limited both from experimental uncertainties and intrinsic
quantum bounds. Here we present an experiment where the amount of
entanglement of a family of two-qubit mixed photon states is
estimated with the ultimate precision allowed by quantum
mechanics.
\end{abstract}
\maketitle
Entanglement is the central resource of quantum information processing
and the precise characterization of entangled states is a crucial issue
for the development of quantum technologies. In turn, quantification and
detection of entanglement have been extensively investigated, see
\cite{ren1,ren2,ren3} for a review, and different approaches have been
developed to extract the amount of entanglement of a state from a given
set of measurement results \cite{bayE,Wun09,Eis07,KA06}.  Of course, in
order to evaluate the entanglement of a quantum state one may resort to
full quantum state tomography \cite{LNP} which, however, becomes
impractical in higher dimensions and may be affected by large
uncertainty \cite{TE94}.  Other methods, requiring a reduced number of
observables, are based on visibility measurements \cite{Jae93}, Bell'
tests \cite{b1,c2,w3},  entanglement witnesses \cite{h1,t2,G3} or are
related to Schmidt number \cite{pas,fed}. Many of them found an
experimental application \cite{mg,wei,ser,nos}, also in the presence of
decoherence effects \cite{buc06,dav07}. 
\par 
Any quantitative measure of entanglement corresponds to a nonlinear
function of the density operator and thus cannot be associated to a
quantum observable. As a consequence, ultimate bounds to the precision
of entanglement measurements cannot be inferred from uncertainty
relations. Any procedure aimed to evaluate the amount of entanglement of
a quantum state is ultimately a parameter estimation problem, where the
value of entanglement is indirectly inferred from the measurement of one
or more proper observables \cite{EE08}. An optimization problem thus
naturally arises, which may be properly addressed in the framework of
quantum estimation theory \cite{qet1,qet2}, which provides analytical
tools to find the optimal measurement and to derive ultimate bounds to
the precision of entanglement estimation.
\par
Suppose one has a family of quantum states $\varrho_\epsilon$
labeled by the value of entanglement, say negativity, and wants to
estimate $\epsilon$ from the outcomes $\chi=(x_1,..,x_N)$ of $N$
repeated measurements of the (generalized) observable described by a
Positive Operator Valued Measurement (POVM) $\Pi_x$,
$\sum_x \Pi_x=\openone$. Any inference strategy amounts to find an
{\em estimator}, i.e. a map $\hat \epsilon (\chi)$ from the
experimental sample to the parameter space. According to the
Cramer-Rao theorem the precision of any estimation procedure, i.e.
the variance of any unbiased estimator based on the measurement of
$\Pi_x$ is bounded by the inequality $\hbox{Var} (\hat \epsilon)\geq
[N F_\epsilon]^{-1}$, where $F_\epsilon =\sum_x  p(x\vert
\epsilon)\, [\partial_\epsilon \ln p(x\vert \epsilon)]^2$ is the
Fisher information and $p(x|\epsilon)=
\hbox{Tr}[\varrho_\epsilon\:\Pi_x]$ is the conditional probability
of getting the outcome $x$ when the actual value of entanglement is
$\epsilon$.  Upon maximizing the Fisher information over all the
possible quantum measurements we arrive at the quantum Fisher
information (QFI) $F_\epsilon \leq H_\epsilon=\hbox{Tr}[L_\epsilon^2
\rho_\epsilon]$ expressed in terms of the symmetric logarithmic
derivative $L_\epsilon$, i.e. the selfadjoint operator satisfying
the equation $\partial_\epsilon \rho_\epsilon= \frac12 (L_\epsilon
\rho_\epsilon + \rho_\epsilon L_\epsilon)$. The ultimate bounds to
precision are thus determined by the quantum Cramer-Rao bound (QCRB)
$\hbox{Var} (\hat \epsilon) \geq [N F_\epsilon]^{-1} \geq [N
H_\epsilon]^{-1}$.  The meaning of QCRB is that quantum mechanics
does not allow entanglement estimation with arbitrary precision. In
turn, QCRB represents the ultimate bound to the precision of {\em
any} procedure aimed to estimate the amount of entanglement of a
state belonging to the family $\varrho_\epsilon$. In order to
optimally estimate entanglement we thus need i) a measurement with
Fisher information $F_\epsilon=H_\epsilon$ equal to the QFI and ii)
an estimator saturating the Cramer-Rao bound \cite{LQE}.  In
\cite{EE08}, bounds to precision have been evaluated for several
classes of pure and mixed quantum states. 
Here we demonstrate experimentally for the first time that optimized
correlations measurements allows for the estimation of entanglement with the
ultimate precision imposed by quantum mechanics.  In particular, we
present the results of an experiment to estimate the amount of
entanglement (negativity) of two-qubit photon states.  This represents a
substantial advance, paving the way for further progresses. In fact,
with a judicious choice of correlation measurements one can devise a
procedure to optimally estimate entanglement for a generic class of
two-photon entangled states. 
\par
\begin{figure}[h]
\includegraphics[width=0.95\columnwidth]{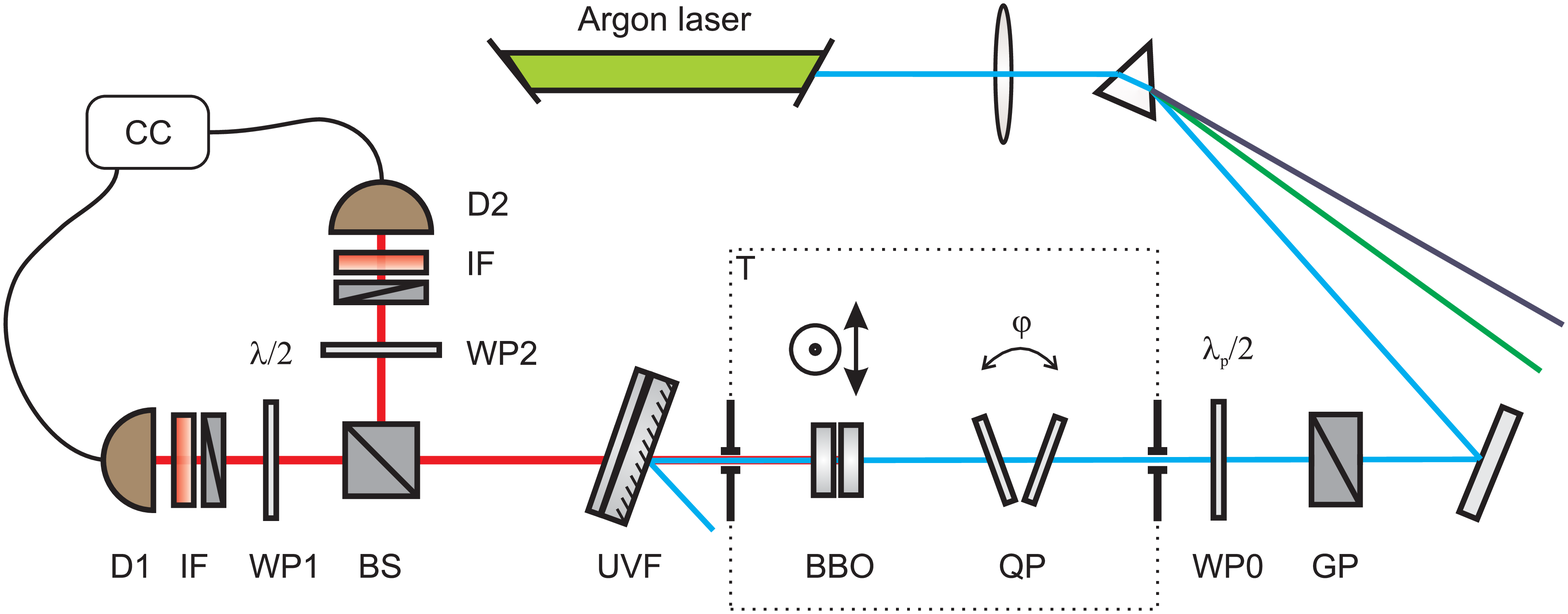}
\caption{(Color online) Experimental setup to generate polarization
entangled photon pairs with variable entanglement and estimate its value
with the ultimate precision allowed by quantum mechanics.}
\label{f:setup}
\end{figure}

The family of entangled states we are dealing with is made of
polarization entangled photon pairs obtained by coherently superimposed
type-I parametric downconversion (PDC) generated in two BBO crystals
\cite{mg}. The experimental set-up is schematically depicted in Fig.
\ref{f:setup}.
A CW Argon pump laser beam with wavelength $\lambda=351nm$ is
filtered with a dispersion prism and then passes through a Glan-Thompson
prism with horizontal axis of transmission that selects the polarization
$\ket{H}$. An halfwave plate WP0 rotates the polarization by the angle
$\phi$, which in turn determines the amount of entanglement in the
output state. PDC light is generated by two thin type-I BBO crystals
(l=1mm), positioned with the planes that contain optical axes
orthogonal to each other. PDC occurs only in crystal 1(2) if the
polarization of the pump beam is horizontal(vertical).
\par
The crystals are cut for collinear frequency degenerate phase-matching
at working wavelength and the phase shifts due to ordinary and extraordinary
path in the crystals are compensated by rotating the quartz plates QP.
To maintain stable phase-matching conditions, BBO crystals and
 QP are placed in a closed box which is kept heated at
fixed temperature. Overall, the output states are described by the
family of density matrices $\varrho_\epsilon = p
|\psi_\phi\rangle\langle\psi_\phi | + (1-p) D_\phi$ which represent
the mixing of the ideal pure entangled states $|\psi_\phi\rangle =
\cos\phi |HH\rangle + \sin\phi|VV\rangle$ (negativity $\epsilon=\sin
2\phi$) with a small fraction of a separable mixture $D_\phi = \cos^2
\phi |HH\rangle \langle HH | + \sin^2\phi|VV\rangle\langle VV|$. The
angle $\phi$ may be tuned at will upon rotating the waveplate WP0,
whereas the mixing parameter $p$ comes from the decoherence
mechanisms occurring in the experimental setup. These are mostly due
to fluctuations of the relative phase between the two polarization
components, which themselves derive from residual
temperature fluctuations: in order to obtain different mixed states
these fluctuation are eventually magnified by eliminating the
isolating box. The entanglement (negativity) of the state
$\varrho_\epsilon$ is given by $\epsilon=p\sin{2\phi}$ and its
purity by $\mu=1-(1-p^2)\sin{2\phi}$. Upon inverting these relations
and express the family of states in terms of $\epsilon$ and $\mu$
or $p$ we may evaluate the QFI for entanglement estimation, which
turns out to be a function of $\epsilon$ only,
$H_\epsilon=(1-\epsilon^2)^{-1}$. In the following, we describe our
detection strategy and show it allows entanglement estimation with
precision saturating the QCRB independently on the purity.
\par
After the crystal, the pump is stopped by a
filter (UVF), and the biphoton field is split on a non-polarizing 50-50
beam splitter (BS). The measurement stage consists in projecting the
beams on vertical polarizers after passing through halfwave plates (WP1,
WP2). After spectral selection by an interference filters (IF) centered at the
degeneracy 702nm (FWHM=3nm), biphotons are collimated by a short
focal length lens and detected on SPAD (D1, D2). Electrical
signals from detectors are registered with a coincidence scheme (CC)
with a time window of coincidences set to $1ns$. Overall, the
measurement scheme is described by projection measurements onto
two-qubit states
$$
\Pi_t (\alpha,\beta)=
\ket{\alpha + s \frac{\pi}2}
\bra{\alpha + s \frac{\pi}2}\otimes
\ket{\beta+ s' \frac{\pi}2}
\bra{\beta+ s' \frac{\pi}2}
$$
where $t=\{s+2s'\}$, $s,s'=0,1$. The polarization angles
$\alpha,\beta$ are set by Glen-Thompson polarizers, whereas the
rotations of $\pm \pi/2$ are obtained by putting the half-wave
plates WP1 and WP2 at $\pm22,5^o$ (mounted on precision rotation
stages with high resolution and fully motor controlled).
\par
Let us first illustrate precision analysis assuming the generation
of the pure states $|\psi_\phi\rangle$. In this case negativity is
given by $\epsilon=\sin2\phi$ and the estimation of entanglement
reduces to a measurement of coincidence rates in a two-particle interferometer
setting \cite{Jae93}. Indeed, upon inspecting the expression of the
probabilities $p_t(\epsilon;\alpha,\beta) =\langle\psi_\phi|
\Pi_t(\alpha,\beta)|\psi_\phi\rangle$, $t=0,1,2,3$ one finds out
that an unbiased estimator for the negativity can be written as
$\hat\epsilon = V(\alpha,\beta) \csc{2\alpha}\csc{2\beta} - \cot
(2\alpha) \cot (2\beta)$, where $V(\alpha,\beta)=p_0-p_1-p_2+p_3$ is
the expected value of two-qubit quantum correlations (QC). 
The corresponding Fisher
information is thus given by $F_\epsilon = \sum_t
p_t(\epsilon;\alpha,\beta)[\partial_\epsilon\log
p_t(\epsilon;\alpha, \beta)]^2$ and it equals the QFI, $H_\epsilon$,
for $\alpha=\pm\pi/4$, $\beta=\pm\pi/4$. In other words
$\hat\epsilon=V(\pm\pi/4,\pm\pi/4)$, which can be measured with the
experimental setup described above, are good candidates for being
optimal estimators of entanglement. In practice, at fixed $\phi$,
$\alpha$ and $\beta$ and for each measurement run $j=1,..,M$ one
records the vector
$\mathbf{k}_j=\{k_{0,j},k_{1,j},k_{2,j},k_{3,j}\}$, where
$k_{t,j}\equiv k_{t,j}(\alpha,\beta)$, is the number of coincidence
counts, as measured for the given set of parameters by the
coincidence circuit during a single time window ($10$ seconds). For
large values of the total number of  coincidences $K_j=\sum_t
k_{t,j}$ (determined irrespectively of the polarizers' orientation
in the $j-th$ run), the expected value of the coincidence rate
$k_{t,j}(\alpha,\beta)/K_j$ converges to the probability
$p_t(\epsilon;\alpha,\beta)$ and the estimator can thus be written
in terms of the coincidences' vector $\hat\epsilon \equiv
\hat\epsilon(\bf{k}_j)$. In our implementation we have performed
$M=30$ measurements of the coincidence vector for fixed values
$\alpha=-\pi/4$ and $\beta=\pi/4$.
\par
For finite $K_j$s the uncertainty in the estimation of entanglement
are mostly due to fluctuations $\delta k_t$ in the coincidence
counts $k_{t,j}$ around their average values $\average{k_t}=\sum_j
k_{t,j}/M$. Using standard error propagation with the derivatives
$\partial_t \equiv \partial/\partial k_t$ evaluated for $k_t\equiv
\average{k_t}$, and assuming independence among fluctuations at
different angles, we have $\hbox{Var} (\hat\epsilon) =
\sum_t |\partial_t\hat\epsilon|^2 \delta k_t^2
=4[(\average{k_0}+\average{k_3})^2(\delta k_1^2+\delta
k_2^2)+(\average{k_1} +\average{k_2})^2(\delta k_0^2+\delta
k_3^2)]/\average{K}^4$. If we now assume that the counting
processes have a Poissonian statistics, i.e. $\delta k_t^2=
\average{k_t}^2$, then it is straightforward to prove that
$$\hbox{Var}(\hat\epsilon) =4
(k_0+k_3)(k_1+k_2)/\average{K}^3 = (1-\hat\epsilon^2)/\average{K}$$ 
i.e. QC measurements allow for optimal estimation of
entanglement with precision at the quantum limit. Since the QFI is
given by $H_\epsilon=(1-\epsilon^2)^{-1}$ for a wide range of
two-qubit families of states \cite{EE08}, the above calculations
suggest that this is a general result. In other words, given a
source emitting polarization two-qubit states with coincidence
counting statistics satisfying the Poissonian hypothesis, then the
experimental setup of Fig \ref{f:setup} allows for optimal
estimation of entanglement at the quantum limit by means of a
QC estimator.
\par
This is indeed the case for the experimental setup of Fig.
\ref{f:setup} and the family of output states $\varrho_\epsilon$
where, upon evaluating the probabilities
$p_t(\epsilon;\alpha,\beta)=\hbox{Tr}[
\varrho_\epsilon\:\Pi_t(\alpha,\beta)]$, one sees that
$\hat\epsilon=V(-\pi/4,\pi/4)$ is still an optimal (unbiased)
estimator of entanglement. In practice, we have collected $M=30$
repeated acquisitions of coincidence vector $\boldsymbol{k}_j =
\{k_{0j},k_{1j},k_{2j},k_{3j}\}$, then we have randomized the
composition of $\boldsymbol{k}_j$ over the sequence of measurements
to avoid spurious correlations, and finally we have estimated
entanglement as the sample mean $\langle\hat\epsilon\rangle = \sum_j
\hat\epsilon(\boldsymbol{k}_j)/M$. The corresponding uncertainty has
been evaluated by the sample variance $\hbox{Var}(\hat\epsilon) =
\sum_j
[\hat\epsilon(\boldsymbol{k}_j)-\langle\hat\epsilon\rangle]^2/(M-1)$.
In order to compare the estimated value of entanglement with the
actual one we need to estimate also the additional parameter $p$,
quantifying the amount of mixing introduced by decoherence
processes. An unbiased estimator $\hat p$ for this parameter may be
obtained by measuring QC with a
different set of angles, e.g. upon collecting the coincidences
$\bf{r}_{j}\equiv \bf{k}_{j}(\alpha=\beta=0)$ to form $\hat p
(\boldsymbol{r}_j,\boldsymbol{k}_j)=
\frac12\hat\epsilon(\boldsymbol{k}_j)
R_j/\sqrt{r_{3,j}(1-r_{3,j})}$, where $R_j=\sum_t r_{t,j}$ is the
total number of coincidences with the four orientations
$\alpha,\beta=0,\pi/2$. The actual ("true") value of negativity is
then inferred as $\epsilon_t=\langle \hat p\rangle\:\sin{2\phi}$,
i.e. using the knowledge of the rotation angle of the waveplate WP0
and the estimation of the mixing parameter.
\begin{figure}[h!]
\includegraphics[width=0.9\columnwidth]{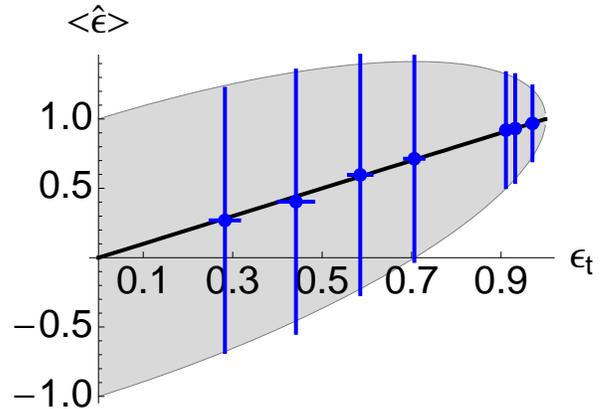}
\caption{Estimation of entanglement at the quantum limit.
The plot shows the estimated value of entanglement $\langle\hat\epsilon\rangle$
as a function of the actual one $\epsilon_t$.
The uncertainty bars on $\langle\hat\epsilon\rangle$ denotes
the quantity $\sqrt{\hbox{Var}(\hat\epsilon) \times \langle K\rangle}$,
i.e. the square root of the sample variance multiplied by the
average number of total coincidences $\langle K \rangle$.
The gray area corresponds to values within the inverse
of the Fisher information $\epsilon_t\pm H_{\epsilon_t}^{-1/2}$.
Uncertainty bars on the
abscissae correspond to fluctuations
$\delta\epsilon_t=\sqrt{\hbox{Var}(\hat p)}\sin2\phi $
in the determination of $\epsilon_t$, due to fluctuations
in the estimation of the mixing parameter.}
\label{f:VarQFI}
\end{figure}
\par
\begin{figure}[h!]
\includegraphics[width=0.45\columnwidth]{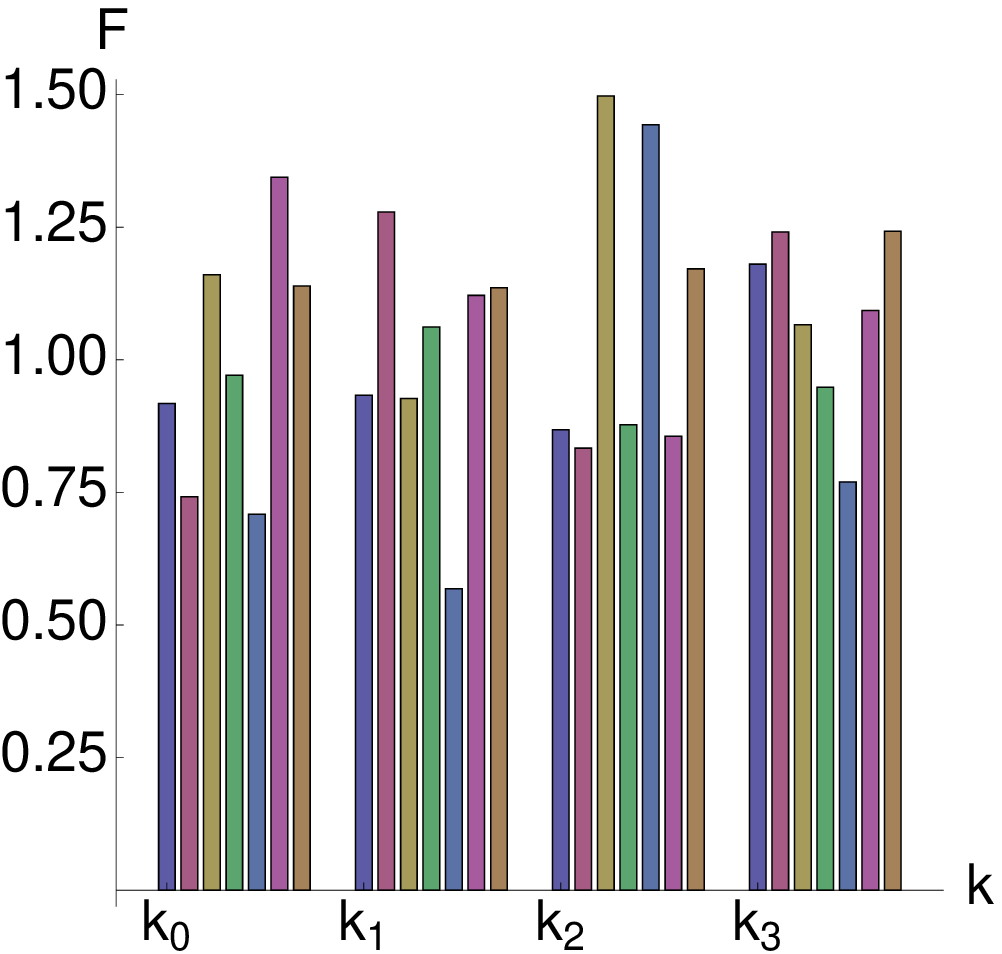}
\includegraphics[width=0.49\columnwidth]{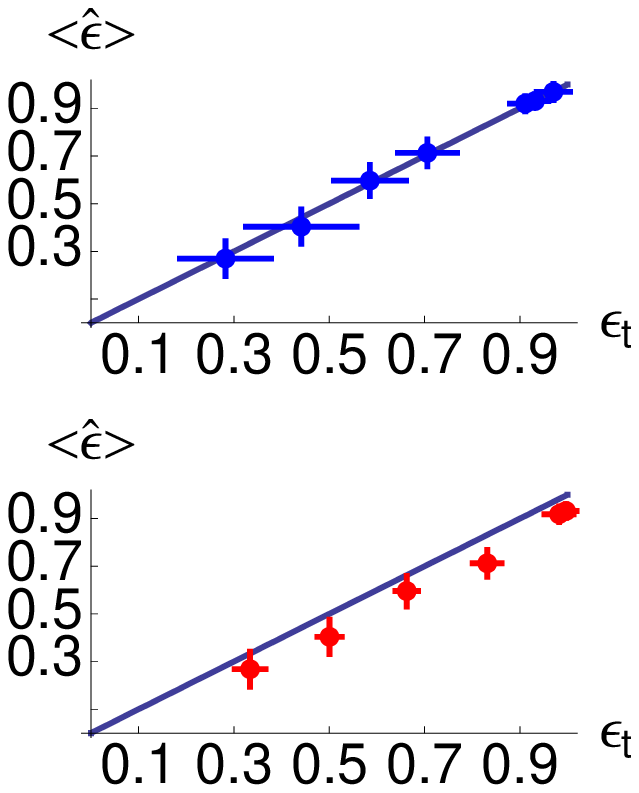}
\caption{(Color online) Left: Fano factors of the coincidence counts
$k_j$, $j=0,1,2,3$. Each groups contains the Fano factor for the seven
values of $\phi$ reported in the text. Right: estimated value of entanglement
as a function of the actual one assuming, for the description of the
output signals, the families
$\varrho_\epsilon$ (top plot) and $\varrho_\epsilon^\prime$ (bottom plot).
The uncertainty bars stays for to the $3\sigma$ confidence
interval.}
\label{f:2}
\end{figure}

In Fig. \ref{f:VarQFI} we show the estimated value of entanglement
as a function of the actual one for seven values of 
$\phi=10^\circ, 15^\circ, 20^\circ, 28^\circ, 40^\circ, 45^\circ, 45^\circ$
of the WP0 rotation angle (corresponding to estimated mixing 
$\langle \hat p\rangle= 0.85, 0.88, 0.88, 0.85, 0.92, 0.93, 0.97$ 
respectively). 
The uncertainty bars on $\langle\hat\epsilon\rangle$ denotes
the quantity $\sqrt{\hbox{Var}(\hat\epsilon) \times \langle K\rangle}$,
i.e. the square root of the sample variance multiplied by the
average number of total coincidences $\langle K \rangle$.
This is in order to allow
a direct comparison with the Cramer-Rao bound in term of the inverse of the
Fisher information (the gray area). Uncertainty bars on the
abscissae correspond to fluctuations
$\delta\epsilon_t=\sqrt{\hbox{Var}(\hat p)}\sin2\phi $
in the determination of $\epsilon_t$, due to fluctuations
in the estimation of the mixing parameter.
As it is apparent from the plot entanglement is estimated with precision
at the quantum limit for any value of the the rotation angle $\phi$.
Notice that this conclusion is robust against the fact that the statistics is
not exactly Poissonian: in the left panel of Fig. \ref{f:2} we
show the Fano factor for the four recorded coincidence counts $k_j$ and
the six values of $\phi$  employed in the experiment.
\par
Our statistical model $\varrho_\epsilon$ may be checked for consistency
on the basis of the recorded data themselves, and other possible models
to describe decoherence of our family of states are ruled out as they
cannot fit the experimental sample. As for example, if one tries to describe
the output from our source by the family of (depolarized) Werner states
$\varrho_\epsilon^\prime = p |\psi_\phi\rangle\langle\psi_\phi | +
\frac14 (1-p)
\openone\otimes\openone$, than one sees from the expression of the 
coincidence probability $p_t^\prime(\epsilon; \alpha,\beta)=\hbox{Tr}[\varrho_\epsilon^\prime\:
\Pi_t(\alpha,\beta)]$
that unbiased estimators for the mixing parameters and the negativity
may be expressed as $\hat p^\prime = V(0,0)$, $\hat \epsilon^\prime
= -\frac12 + \frac12 V(0,0) + V(-\pi/4,\pi/4)$. These may be written 
in terms of the coincidence vectors $\mathbf{k}$ and
$\mathbf{r}$ as
$\hat p^\prime (\boldsymbol{r}_j)=(r_{r0,j}-r_{1,j}-r_{2,j}+r_{3,j})/R_j$ and
$\hat\epsilon^\prime (\boldsymbol{r}_j,\boldsymbol{k}_j)= -\frac12 +
\frac12 \hat
p^\prime (\boldsymbol{r}_j)+ (k_{0,j}-k_{1,j}-k_{2,j}+k_{3,j})/K_{j}$.
Upon evaluating the corresponding sample means and variances one
realizes that the model is incompatible with the observed data. This is
illustrated in the right panel of Fig.\ref{f:2} where we report
the estimated value of entanglement as a function of the actual one
assuming, for the description of the output signals, the families
$\varrho_\epsilon$ (top plot) and $\varrho_\epsilon^\prime$ (bottom plot).
Here the uncertainty bars denote the $3\sigma$ confidence
interval and thus it is apparent that $\varrho_\epsilon^\prime$ cannot
fit the data.
We also performed measurements of the coincidence counts
for a complete set of polarization angles and check the
model $\varrho_\epsilon$ by full two-qubit state
tomography \cite{KB00,Ja01}.
\par
\par
In conclusion, we have suggested and demonstrated a measurement scheme
based on quantum correlation measurements to optimally estimate
entanglement for a family of two-photon entangled states.  Our procedure
is self-consistent and allows to estimate the amount of entanglement
with the ultimate precision imposed by quantum mechanics.  With an
appropriate choice of correlation measurements our results may be
extended to a generic class of two-photon entangled states. The
statistical reliability of our method suggests a wider use in precise
monitoring of external parameters assisted by entanglement.
\par
This work has been supported in part by MIUR (PRIN 2007FYETBY),
Regione Piemonte (E14),  "San Paolo foundation", and NATO
(CBP.NR.NRCL.983251). MGAP and PG thanks Marco Genoni for several 
useful discussions.

\end{document}